\documentclass{article}

\begin{document}

\title{Burgers' equation in non-commutative space-time}
\author{L. Martina $^{*}$ and O. K. Pashaev $^{\dagger}$ \\
 $^{*}$ Dipartimento di Fisica and Sezione INFN - Lecce, \\
Universita di Lecce,
Lecce, 11529, Italy\\
 $^{\dagger}$ Department of Mathematics,
Izmir Institute of Technology, \\
 Urla-Izmir, 35437, Turkey}
\maketitle \abstract{The Moyal $*$-deformed noncommutative
version of Burgers' equation is considered. Using the $*$-analog
of the Cole-Hopf transformation, the linearization of the model
in terms of the linear heat equation is found. Noncommutative
q-deformations of shock soliton  solutions and their interaction
are described  }
\section{Introduction}
In the last decades the idea of a noncommutative field theory have
received a great interest in the context of the string theory
\cite{Seiberg}, in connection with the Noncommutative Geometry,
created by A. Connes and many others \cite{Connes},
\cite{Varilly1}. \ It has been applied to the Standard model in
the unified field theory \cite{Varilly2} \,    gravity theory
\cite {Chamseddine} and the quantum Hall systems. In this context
several attempts to extend the completely integrable, or
solitonic, classical field theories to non commutative spaces have
appeared \cite{Dimakis}. In the last case the basic idea is to
include into the characterizing structures of the completely
integrable systems ( the bicomplex, or in a more traditional way,
the Lax pair) the non commutative nature
 of the space-time. Then, single soliton - like solutions for the noncommutative
analogs of the Nonlinear Schr{\"o}dinger and Korteweg-de Vries
equations (NLS and  KdV, respectively) were presented, in terms of
power expansion in the deformation parameter $\theta$. However,
the actual problem constructing of two and higher soliton
solutions and studying  the corresponding dynamics is faced with
difficulties. In the present paper we introduce the noncommutative
version of the Burger's equation, which is linearizable via the
*-analog of Cole - Hopf transformation, and construct two-shockes
soliton solution in terms of q-deformed binomial series. This
allows us to analytically deal with the two shocks fusion process
into a final one solitonic object.
\section{The $*$ - product}
The Moyal $*$-product\cite{moyal} is an associative and non commutative
deformation of the ordinary product between two functions $f(x_1, x_2)$, $%
g(x_1, x_2)$ $\in C^\infty$ on $R^2$
\begin{equation}
\begin{array}{rcl}
&  & (f*g)(\mathbf{x}) = e^{\theta (\partial_1 \partial_{2^{\prime}} -
\partial_2 \partial_{1^{\prime}})}f(x_1, x_2)g(x_1^{\prime}, x_2^{\prime})|_{%
\mathbf{x} = \mathbf{x^{\prime}}} \\[4pt]
&  & = \sum_{n= 0}^{\infty} \frac{\theta^n}{n!}\sum_{k= 0}^{n} (-1)^k \left(%
\matrix{n\cr k\cr}\right)(\partial^{n-k}_1 \partial^k_2 f)(\partial^k_1
\partial^{n-k}_2 g)\,\,, \\
&  &
\end{array}
\label{1}
\end{equation}
where $\mathbf{x} = (x_1, x_2) \in R^2$. The parameter $\theta$
is assumed to be purely immaginary, but for our aims, we consider
also the analytic continuations to the real $\theta$. It is known
that the convergence of the series in (\ref{1}) requires that the
product be defined in a space of rapidly decaying functions
(usually of Schwartz type) on the plane. With such a definition
one easily sees that $\lim_{\theta \rightarrow 0} f * g = f g $.
If (\ref{1}) converges also for real values of $\theta$ the
resulting function is real, but in general is not rapidly
decaying, thus one needs some care to deal with. Moreover, if the
series (\ref {1}) converges, one can use the integral
representation
\begin{equation}
f\ast g\left( \mathbf{\vec{x}}\right) =-\frac{1}{\left( \pi \theta \right)
^{2}}\int \int d^{2}x^{\prime }d^{2}x^{^{\prime \prime }}f\left( \mathbf{%
\vec{x}}^{\prime }\right) g\left( \mathbf{\vec{x}}^{\prime \prime }\right)
e^{-\frac{1}{\theta } \left( \mathbf{\vec{x}}^{\prime }-\mathbf{\vec{x}}%
\right) \times \left( \mathbf{\vec{x}}^{\prime \prime }-\mathbf{\vec{x}}%
\right) }.  \label{Bakerform}
\end{equation}
The basic example of $*$-product is between expontials:
\begin{equation}
e^{\mathbf{{k}\cdot {x}}}*e^{\mathbf{{\ k^{\prime }}\cdot {x}}}=e^{\theta
\mathbf{{k }\times {k^{\prime }}}} e^{\left( \mathbf{{k}+ {k^{\prime }}}%
\right) \cdot {x}}.  \label{eq:com}
\end{equation}
The $*$-product of two gaussian functions provides again an exponential with
quadratic argument, but its coefficient and the amplitude in front of the $%
exp$ are depending from $\left( 1-ab\theta^2\right)^{-1}$. Thus,
one sees that for $\theta$ immaginary the $*$-product spreads the
packets,  while for arbitrary real values the resulting function
can be unbounded.

The $*$- product enables us to define the Moyal Brackets as $\{f,
g\}_{\theta} \equiv \left(f * g - g * f\right)/(2 \theta)$, which
realize a deformation of the Poisson brackets, in the sense that
they provide a Lie algebra on the functions of phase space
$\left( x_1, x_2 \right)$ and the relation $\lim_{\theta
\rightarrow 0} \{f, g\}_{\theta} = \{f, g\}_{Poisson} $ holds.

Since the $*$-product involves exponentials of the derivative operators, it
may be evaluated through
\begin{equation}
(f * g)(\mathbf{x}) = f(\hat x_1, \hat x_2) g(\mathbf{x}),  \label{eq:oper2}
\end{equation}
where the uncertainty relation holds for the position operators
\begin{equation}
\hat x_i = x_i + \theta \epsilon_{i j} \partial_j, \qquad [\hat x_1, \hat
x_2] = 2\theta\,\,.  \label{eq:com1}
\end{equation}
Thus, the algebra of function on $\mathbf{R^2}$ endowed with the $*$-product
is describing the noncommutative plane.

\section{Noncommutative Burgers' equation}

\subsection{The linear problem}

In the noncommutative space-time ($x_1 = t$, $x_2 = x$) plane we consider
the linear problem
\begin{equation}
\chi_x = U \,*\,\chi\,\,,\,\,\,\,\,\chi_t = V \,*\,\chi\,\,,
\label{eq:lin1}
\end{equation}
with the noncommutative Abelian connections
\begin{equation}
U = -\frac{1}{2\nu}\,u\,\,, \,\,\,V = -\frac{1}{2\nu}(\nu u_x - \frac{1}{2}
u*u)\,\,.  \label{eq:pot}
\end{equation}

The translations operators $\partial_t$ and $\partial_x$ are
derivations w.r.t. the $*$-product. Then, the integrablility of
(\ref{eq:lin1}) is assured by the usual equality of the mixed
derivatives. Consequently, the zero curvature equation in Moyal
form
\begin{equation}
U_t - V_x + U * V - V * U = U_t - V_x + 2\theta\{ U , V\}_\theta = 0\,\,,
\label{eq:zc}
\end{equation}
provides the Noncommutative Burgers' (NB) equation
\begin{equation}
u_t + u_x \,*\, u = \nu\, u_{xx}.  \label{eq:NB}
\end{equation}
The above noncommutative flat connection representation shows the
relevance of our model for the BF topological field theory and
noncommutative low dimensional gauge theory of gravity\cite{cham}.

Given a solution $\chi$ of this problem, we define first the $*$-inverse function $%
_{*}\chi^{-1}$:
\begin{equation}
\chi \, * \,_{*}\chi^{-1} = 1\,\,,  \label{eq:inv}
\end{equation}
possessing properties: a) $_{*}\chi^{-1} = \chi^{-1}_*$, b) $%
(_{*}\chi^{-1})_x = -_{*}\chi^{-1}\,* \,\chi \,*\, _{*}\chi^{-1}$. In
general the $*$-inverse, if it exists, is not unique. But, since we are
going to deal with exponentials and finite sums of exponentials, the
relation (\ref{eq:com}) assures existence and uniqueness.  Then, inverting
the first of Eqs.(\ref{eq:lin1}) we get the $*$-analog of the Cole-Hopf
transformation
\begin{equation}
u = (- 2\nu) \, \chi_x \,*\,_{*}\chi^{-1}\,\,  \label{eq:CH}
\end{equation}
relating a solution of the heat equation
\begin{equation}
\chi_t = \nu \,\chi_{xx}\,\,,  \label{eq:heat}
\end{equation}
to a solution of the NB equation.
Furthermore,  putting $_{*}\chi^{-1} \equiv v$, we have
another noncommutative nonlinear heat equation
\begin{equation}
v_t + 2\nu \, v_x \,*\,_{*}v^{-1}\,*\,v_x =   \nu \,v_{xx}.
\label{eq:NH}\end{equation}

\subsection{Noncommutative shock soliton}

Now we consider particular solutions of the heat equation (\ref{eq:heat}) in
the exponential form $e^{\eta_j}$, $\eta_j = \omega_j t + k_j x +
\eta^{(0)}_j$, $\omega_j = \nu k^2_j$, and their superpositions, generating
by transformation (\ref{eq:CH}) exact solutions of the NB (9). Again from
(3) we get
\begin{equation}
e^{\eta_i} \,*\, e^{\eta_j} = e^{\theta \Delta_{ij}} e^{\eta_i + \eta_j} =
e^{2\theta \Delta_{ij}}e^{\eta_j} \,*\, e^{\eta_i},  \label{eq:qcom}
\end{equation}
where $\Delta_{ij} = - \Delta_{ji} = \omega_i k_j - \omega_j k_i = \nu k_i
k_j (k_i - k_j)$.

For $\chi = e^{\eta_1}$, the $*$-inverse function is simply $_* \chi^{-1} =
e^{-\eta_1}$, and from Eq. (\ref{eq:CH}) we obtain the constant valued
solution of Eq.(9): $u(t,x) = (-2\nu)k_1 $.

Taking a superposition of two exponential solutions $\chi = e^{\eta_1} +
e^{\eta_2}$ , first we look for the $*$-inverse as a power series $_*
\chi^{-1} = \sum^{\infty}_{n= 0} \epsilon_n \,e^{\Omega_n t + P_n x + C_n
}\,\, \label{eq:inv1}$. Equation (\ref{eq:inv}) allows to find $\epsilon_n =
(-1)^n$, $\Omega_n = -\omega_1 + n (\omega_2 - \omega_1)$, $P_n = - k_1 + n
(k_2 - k_1) $, $C_n = - \eta^{(0)}_1 + n (\eta^{(0)}_2 - \eta^{(0)}_1)$.
This series is convergent in the two regions $e^{\eta^-_{21}} < 1$ and $%
e^{\eta^-_{12}} < 1$, where $\eta^-_{ij} \equiv \eta_i -\eta_j$.  Combining
together the results, for $e^{\eta^-_{12}} \neq 1$ we find that the $*$%
-inverse is
\begin{equation}
_* \chi^{-1} = \frac{1}{e^{\eta_1} + e^{\eta_2}}\,\,.
\end{equation}
By the Cole-Hopf type transformation (\ref{eq:CH}) we are able to find the
1-shock soliton of the NB equation
\begin{equation}
u(t,x) = (-2\nu) \left[k_1 + \frac{k_2 - k_1}{1 + e^{\eta^-_{12} -
\theta\Delta_{12}}}\right]\,\,,  \label{eq:sol}
\end{equation}
with amplitude $k_2 - k_1$ and velocity $v_{12} = -\nu (k_1 +
k_2)$. In this case the noncommutativity influences the initial
position of the single soliton $x^{\theta}_0 = x_0 + \theta \nu
k_1 k_2\,, $ so that $\lim_{\theta \rightarrow 0} x^{\theta}_0
\rightarrow x_0$.

\subsection{Two soliton solution}

Now, let us take as solution of the heat equation (\ref{eq:heat})
$\chi = e^{\eta_1} + e^{\eta_2} + e^{\eta_3} $. For the power
series expansion of its $%
*$-inverse we needs to divide the space-time plane on three
regions and represent it in different ways in that regions.
Specifically, as $_*\chi^{-1} =  e^{-\eta_1} \sum_{ij} a_{ij}
e^{i \eta^-_{21} + j \eta^-_{31}}$ in $D_{1} = \{(x,t):
\eta^-_{21} < 0, \eta^-_{31} < 0\}$, as $_*\chi^{-1} =
e^{-\eta_2} \sum_{ij} a_{ij} e^{i \eta^-_{12} + j \eta^-_{32}}$
in the region $D_{2} = \{(x,t): \eta^-_{12} < 0, \eta^-_{32} <
0\}$ and,  finally, in region $D_{3} = \{(x,t): \eta^-_{13} < 0,
\eta^-_{23} < 0\}$ we look for $_*\chi^{-1} = e^{-\eta_3}
\sum_{ij} a_{ij} e^{i \eta^-_{13} + j \eta^-_{23}}$. Substituting
these expessions into the Eq.(\ref{eq:inv}),  we obtain that
$a_{ij} = (-1)^{i+j} b_{ij}$, where coefficients $b_{ij}$ satisfy
the recurrence relations
\begin{equation}
b_{ij} = b_{i-1\,j}e^{j\theta \Delta} + b_{i\,j-1}e^{-i\theta \Delta}\,\,,
\label{eq:req}
\end{equation}
and $\Delta \equiv \Delta_{12} + \Delta_{23} + \Delta_{31}$.

\subsection{The q-deformed Pascal-Tartaglia triangle}

The coefficients in (\ref{eq:req}) can be  generated by the q-
deformed version of Pascal-Tartaglia triangle
\[
\matrix{&&&&& b_{00}&&& &&\cr\cr &&&&\swarrow & &\searrow &\cr\cr
& &&b_{10}& &&& b_{01}& &\cr &&\swarrow&  &\searrow  e^{-\alpha}&&
e^{\alpha} \swarrow && \searrow&\cr\cr & b_{20}& &&& b_{11}& &&&
b_{02}&\cr \swarrow & & \searrow e^{-2\alpha}&
&e^{\alpha}\swarrow & & \searrow  e^{-\alpha}& & e^{2\alpha}
\swarrow & & \searrow \cr &\cdots&&&&\cdots&&&&\cdots&\cr}
\]
where $\alpha = \theta \Delta$, $q = e^{\alpha} = e^{\theta \Delta}$. Thus
the $b_{n,k}$ are given by the q- Binomial Coefficients
\begin{equation}
\left(\matrix{n \cr k \cr}\right)_{\alpha} = \frac{[n]!}{[k]![n-k]!}\,\,,
\label{eq:bin}
\end{equation}
constructed from q-numbers $[n]_{\alpha} = \sinh n\alpha$ and the
corresponding q-factorials $[n]! = [1][2]...[n]$. We would like
to notice that "canonical" q-numbers of the form $[n] = \sinh
n\alpha/\alpha$ produce the same q- Binomial coefficients.

Now, we define the generating function (GF) for these coefficients as
\begin{equation}
G_{\alpha }(x,y)=\sum_{ij}(-1)^{i+j}\left( \matrix{i+j \cr j \cr}\right)
_{\alpha }x^{i}y^{j}\,\,.  \label{eq:gen}
\end{equation}
For $\theta =0$ ($\alpha =0$), it reduces to the GF of the ordinary binomial
coefficients $G_{0}(x,y)=\left[ 1+(x+y)\right] ^{-1}$ In terms of GF (\ref
{eq:gen}) we are able to get the formal power series expression for the $%
\ast $-inverse of $\chi $ as
\begin{equation}
_{\ast }\chi ^{-1}=\cases{e^{-\eta_1} G_\alpha (e^{\eta^-_{21}},
e^{\eta^-_{31}})\,\,\,{\rm in} \,\,D_{1}\,\,\,, \cr\cr e^{-\eta_2} G_\alpha
(e^{\eta^-_{12}}, e^{\eta^-_{32}}) \,\,\,{\rm in} \,\,D_{2}\,\,,\cr\cr
e^{-\eta_3} G_\alpha (e^{\eta^-_{13}}, e^{\eta^-_{23}}) \,\,\,{\rm in}
\,\,D_{3}\,.\cr}  \label{eq:vfunc}
\end{equation}
By the $\ast $- Cole-Hopf type transformation (\ref{eq:CH}), it
gives the 2-soliton solution of the NB equation (\ref{eq:NB}),
expressed in power series in the region $D_{i}$,$ (i = 1,2,3)$ by
\begin{equation}
 u_{i}(t,x)=\sum_{j = 1, k<l \neq i}^{3} a_{j}e^{\eta _{ji}^{-}+\theta
\Delta _{ij}}
G_{\alpha }\left(e^{\eta _{ki}^{-}+\theta\left( \Delta
_{ik} + \epsilon_{ijk} \Delta\right)}, e^{\eta_{li}^{-}+\theta\left( \Delta
_{ik} + \epsilon_{ijl} \Delta\right)}\right)
\label{sol1}
\end{equation}
where we have used the following parametrization: $\eta _{ij}^{-}\equiv \eta
_{i}-\eta _{j}=-((a_{i}-a_{j})/2\nu )[x-v_{ij}t-{x_{0}}_{ij}]=k_{ij}^{-}\xi
_{ij}$, $v_{ij}=(a_{i}+a_{j})/2$, ${x_{0}}_{ij}=(2\nu /(a_{i}-a_{j}))(\eta
_{i}^{(0)}-\eta _{j}^{(0)})$. Then, for $a_{3}>a_{2}>a_{1}$ we have $%
k_{12}^{-}>0,\,k_{13}^{-}>0,\,k_{23}^{-}>0$ and
$v_{23}>v_{13}>v_{12}$ for the velocities. Such velocities
determine the motion in $(x,t)$ plane with the boundary lines
defining the domains $D_{1}$, $D_{2}$ and $D_{3}$, respectively.
In the asymptotic region of these domains one of the arguments of
GF (\ref{eq:gen}) vanishes, so that GF reduces to $G_{\alpha
}(0,y)=G_{\alpha }(y,0)=\left( 1+y\right) ^{-1}$ , for $y<1$.
Then, we find that for any fixed time $t_{0}$ the solution decays
exponentially at infinities
\begin{equation}
u(t_{0},x)=\cases{a_3\,\,, \hskip1cm x \rightarrow -\infty\cr a_1\,\,,
\hskip1cm x \rightarrow +\infty\cr}.  \label{eq:asymp}
\end{equation}
On the other hand, one can study the asymptotic behaviour of the solution in
each of the moving frame of relative velocity $v_{ij}$ above defined.
We summarize in the following table the relative velocity, the
limiting points and the asymptotic behaviour of the solution
\[
\begin{array}{ccc}
v_{12},x\rightarrow -\infty ,t\rightarrow -\infty ; & v_{23},x\rightarrow
-\infty ,t\rightarrow -\infty  ; & v_{13},x\rightarrow +\infty ,t\rightarrow
+\infty , \\
u\rightarrow \frac{a_{1}+a_{2}e^{\eta _{21}^{-}+\theta \Delta _{12}}}{%
1+e^{\eta _{21}^{-}+\theta \Delta _{12}}}\,\,, & u\rightarrow \frac{%
a_{2}+a_{3}e^{\eta _{32}^{-}+\theta \Delta _{23}}}{1+e^{\eta
_{32}^{-}+\theta \Delta _{23}}}\,\,, & u\rightarrow \frac{a_{1}+a_{3}e^{\eta
_{31}^{-}+\theta \Delta _{13}}}{1+e^{\eta _{31}^{-}+\theta \Delta _{13}}}%
\,\,.
\end{array}
\]
Then, it shows that the solution describes a fusion of two isolated shock solitons,
moving with velocities $v_{12}$ and $v_{23}$, into the final one soliton, possessing
velocity $v_{13}$. The qualitative picture of this soliton interaction is
similar to the commutative case. But the "central" positions of these solitons
are modified,  according to the expression ${x_{0}^{\theta }}_{ij}=\frac{2\nu }{%
a_{i}-a_{j}}(\eta _{i}^{(0)}-\eta _{j}^{(0)}-\theta \Delta _{ij})$. Thus, we have the
 relation
\begin{equation}
A_{12}{x_{0}^{\theta }}_{12}+A_{23}{x_{0}^{\theta }}_{23}-2\nu \theta \Delta
=A_{13}{x_{0}^{\theta }}_{13}\,\,,
\end{equation}
which connects the positions ${x_{0}^{\theta }}_{ij}$ and the amplitudes $A_{ij}\equiv
-(a_{i}-a_{j})$ of initial and final solitons.
This formula determines the position of the ''center of mass'' of
the  final soliton, which results  to contain the $\theta $ dependent
shift
\begin{equation}
-2\nu \frac{\theta \Delta }{A_{12}+A_{23}}\,,\label{shift}
\end{equation}
with respect to the commutative ($\theta = 0$) case. Moreover, this shift is
independent on the initial positions of the colliding solitons.
\section{Conclusions}
In this work on the Burger's equation in noncommutative
space-time we have shown how the noncommutativity affects one of
the main feature of the soliton interaction behaviour, i.e. the
shift in position of the final states. In the case of two
interacting shocks Eq. (\ref{shift}) provides an explicit
algebraic expression of that quantity in the case of the process
of fusion of two shocks, linearly dependent on the deformation
parameter $\theta$. The last result can be considered as a test
distinguishing between the commutative and noncommutative
solitons, in the analytical and qualitative studying of soliton's
dynamics. It suggests also that in the real experiments with
optical solitons, appearance of a global position shift
independent of initial soliton  positions indicates on the
noncommutative nature of corresponding solitons.

As is well known the Burgers' equation is an example of the so
called C-integrable solitonic equation, and as we showed in the
present paper the concept of C-integrability can be extended to
the noncommutative space-time case. Moreover it would be
interesting to see the influence of the noncommutativity on the
so called Q-integrable equations, like the NLS and the KdV or the
self-dual Yang-Mills system, as well. The analytic study of the
two- , and possibly multi-, soliton solutions for the
noncommutative analogs of those system is in our plans for
forthcomming future.

 \bigskip

{\bf Acknowledgments}
One of the authors (O.K.P.) would like to thanks Profs. M. Boiti and F.
Pempinelli for invitation and local support. This research is partially
supported by the italian
MIUR and INFN - Sezione di Lecce.

\end{document}